\documentclass[10pt,a4paper,twocolumn]{article}
\usepackage{amsmath,amssymb,mathrsfs,bm}
\usepackage{graphicx,booktabs,array,dcolumn}
\usepackage{url}
\usepackage{microtype}
\usepackage[margin=1.9cm,columnsep=0.65cm]{geometry}
\begin{document}
\title{Primordial non-Gaussianity in warm $k$-inflation with nonminimal derivative coupling}
\author{Xiao-Min Zhang \quad Zi-Xin Bai \quad Qing-Ze Shi\\
Run-Qing Zhao \quad Peng-Cheng Chu \quad Yun-Cai Feng\\[0.55em]
\small School of Science, Qingdao University of Technology, Qingdao 266520, China\\
\small Corresponding author: \texttt{zhangxm@mail.bnu.edu.cn}}
\date{}
\maketitle
\begin{abstract}
This paper derives the finite-inertia intrinsic thermal bispectrum in potential-free warm $k$-inflation with nonminimal derivative coupling (NMDC). On a fixed Friedmann--Lema\^{\i}tre--Robertson--Walker geometry, the complete cubic inflaton action determines the nonlinear stochastic source, encompassing temporal, spatial-gradient, and field-dependent kinetic interactions. The retarded solution captures the full momentum-triangle dependence while retaining unequal-time correlations, whereas its equilateral limit yields the intrinsic nonlinearity parameter. The power spectrum is computed with the same response and Gaussian initial state, thereby fixing the statistical normalization. A complementary local freeze-out reduction makes the interaction coefficients and two geometric shape contributions explicit. Comparison along an evolving near-attractor trajectory indicates that unequal-time finite-inertia propagation is crucial for quantitatively predicting the intrinsic bispectrum. The analysis also derives the slow-roll-suppressed nonlinear $\delta N$ conversion, together with the Arnowitt--Deser--Misner constraints and cubic vertices that characterize the direct gravitational effects of NMDC. These results clarify the respective roles of local freeze-out estimates and finite-inertia stochastic evolution in the nonlinear prediction.
\end{abstract}

\section{Introduction}

Inflation was proposed to resolve the horizon, flatness, and monopole problems of the standard Big Bang model \cite{Guth:1981,Linde:1982,Albrecht:1982,Bassett:2006}. It also provides a mechanism for generating primordial density perturbations, which subsequently give rise to the observed anisotropies in the cosmic microwave background (CMB) and to large-scale structure \cite{Mukhanov:1992,Liddle:2000}.

In cold inflation, radiation is generally produced only after inflation has ended. By contrast, warm inflation \cite{Berera:1995a,Berera:1995b} permits the inflaton to transfer energy to radiation during the inflationary stage. This process can sustain a thermal bath whose fluctuations contribute to the primordial spectrum. Dissipation can also relax the slow-roll requirements~\cite{Berera:2005,Berera:2009}. This is relevant to the inflaton $\eta$-problem associated with supergravity corrections~\cite{Dine:1995,Kolda:1999}. Thermal contributions to the scalar power spectrum have been investigated in Refs.~\cite{Berera:2000,Hall:2004,Graham:2009,Montefalcone:2024}.

In $k$-inflation, the Lagrangian takes a general form $\mathcal{L}(X,\phi)$, allowing accelerated expansion even in the absence of a scalar potential \cite{Armendariz-Picon:1999,Garriga:1999}. Noncanonical kinetic terms modify the propagation speed $c_s$ and can strengthen nonlinear interactions, so primordial non-Gaussianity provides a valuable observational probe \cite{Creminelli:2003,Alishahiha:2004,Silverstein:2004}.

Nonminimal derivative coupling (NMDC) offers a further extension. The interaction $G^{\mu\nu}\partial_\mu\phi\partial_\nu\phi$ enhances gravitational friction and alters the inflaton perturbation equations \cite{Germani:2010,Karydas:2021,Dalianis:2020}. Its consequences have been examined in both cold and warm inflation \cite{Germani:2010,Karydas:2021,Nozari:2014,Sadjadi:2015,Zhang:2025a}.

Our previous work investigated noncanonical warm inflation with NMDC, including models with a potential \cite{Zhang:2025a}. We subsequently studied the potential-free case $\mathcal{L}(X,\phi)=K(\phi)X + X^2/\mu^4$ \cite{Zhang:2026}, analyzing the background dynamics, leading-order scalar spectrum, spectral index and tensor-to-scalar ratio, as well as the Planck inflationary constraints \cite{PlanckInflation:2020}. The three-point function, the subject of the present work, probes interactions beyond those determined by the power spectrum. The bispectrum, often represented through $f_{\mathrm{NL}}$, can reveal interactions that could be difficult to distinguish using the power spectrum alone. In canonical single-field cold inflation, the scalar bispectrum is typically slow-roll suppressed~\cite{Maldacena:2003,Lyth:2005c}. Noncanonical kinetic interactions can generate larger signals~\cite{Chen:2010}, and thermal bispectra have been studied in canonical and noncanonical warm inflation~\cite{Gupta:2002,Moss:2007,Bastero-Gil:2014,Nozari:2014,Zhang:2015,Shen:2023,Zhang:2025b}. Their momentum dependence has also motivated dedicated shape and template analyses~\cite{Mirbabayi:2023}.

In the present work, we extend the non-Gaussianity calculation of warm $k$-inflation without NMDC~\cite{Shen:2023} to the potential-free derivative-coupling model. Our calculation follows the second-order Langevin treatment of thermal fluctuations: inflaton self-interactions produce the intrinsic bispectrum, while nonlinear $\delta N$ conversion contributes the slow-roll component. The full intrinsic quadratic source is derived directly from the cubic inflaton action, including the temporal, spatial-gradient, and field-dependent kinetic vertices. Retaining the inertial term makes it possible to relate this source to the unequal-time thermal bispectrum.

Our main result is the finite-inertia intrinsic bispectrum for general momentum triangles and the corresponding equilateral nonlinearity parameter. First, the two- and three-point functions are evolved from the same Gaussian state, and their common stochastic normalization is fixed. Then, a compact local freeze-out expression disentangles the coupling strengths from the triangle geometry, and it is tested against the retarded result on the same evolving background. Finally, the $\delta N$ result, together with the Arnowitt--Deser--Misner (ADM) constraints and cubic vertices, establishes the accompanying adiabatic conversion and local gravitational interactions.

The rest of this paper is organized as follows. Section~\ref{sec:framework} presents the model and background dynamics. In Sec.~\ref{sec:nonGaussianity}, the $\delta N$ and intrinsic contributions are calculated, and the metric interactions are examined. Section~\ref{sec:results} explores the physical implications, and Sec.~\ref{conclusion} presents our conclusions. Further technical details are given in the appendices.

\section{The framework of warm $k$-inflation with nonminimal derivative coupling}
\label{sec:framework}

First, we recall the potential-free model of Ref.~\cite{Zhang:2026} and its background dynamics.

\subsection{Action and background equations}

The total action of the system is given by
\begin{equation}
\begin{aligned}
S=\int d^4x\,\sqrt{-g}\,\Bigl[&\frac{M_p^2}{2}R
+\mathcal L(X,\phi)\\
&+\frac{G^{\mu\nu}\partial_\mu\phi\partial_\nu\phi}{2M^2}
+\mathcal L_{\mathrm{int}}+\mathcal L_R\Bigr].
\end{aligned}
\label{eq:action}
\end{equation}
Here, $M_p$ denotes the reduced Planck mass, $G^{\mu\nu}$ signifies the Einstein tensor, and $M$ sets the NMDC scale. The inflaton Lagrangian is purely kinetic,
\begin{equation}
\mathcal{L}(X,\phi) = K(\phi) X + \frac{X^2}{\mu^4}, \qquad X = -\frac{1}{2}g^{\mu\nu}\partial_\mu\phi\partial_\nu\phi,
\label{eq:Lagrangian}
\end{equation}
with $K(\phi)<0$ \cite{Zhang:2026}, and the parameter $\overline{K}(\phi)\equiv -K(\phi) > 0$ is defined for later use. The dissipative interactions encoded in $\mathcal{L}_{\rm int}$ give rise to a friction term characterized by $\Gamma$, while $\mathcal{L}_R$ characterizes the radiation sector.

In a flat Friedmann--Lema\^{\i}tre--Robertson--Walker (FLRW) background,
\begin{equation}
ds^2 = -dt^2 + a^2(t)\,\delta_{ij}dx^i dx^j,
\end{equation}
the inflaton obeys
\begin{equation}
\begin{aligned}
&(\mathcal{L}_X c_s^{-2}+3F)\ddot\phi
+3H(\mathcal{L}_X+3F+r)\dot\phi\\
&\qquad+\frac{6H\dot H}{M^2}\dot\phi
+\frac12K_\phi\dot\phi^2=0.
\end{aligned}
\label{eq:motion}
\end{equation}
Here, the following definitions are used
\begin{align}
\mathcal{L}_X \equiv \frac{\partial\mathcal{L}}{\partial X} = K(\phi) + \frac{2X}{\mu^4}, \quad\quad\quad \label{eq:LX} \\
F \equiv \frac{H^2}{M^2}, \qquad
r \equiv \frac{\Gamma}{3H}, \quad\quad\quad\label{Fr} \\
c_s^2 \equiv \frac{p_X}{\rho_X} = \frac{\mathcal{L}_X}{\mathcal{L}_X + 2X\mathcal{L}_{XX}} = \left(1 + \frac{4X/\mu^4}{\mathcal{L}_X}\right)^{-1}. \label{eq:soundspeed}
\end{align}
We retain the term $6H\dot H\dot\phi/M^2$ when evolving the background. The parameter $F$ quantifies the NMDC contribution, while $r$ represents the dissipative ratio. The relative friction enhancement is governed by $3F$ alongside $\mathcal L_X$, while the inertial coefficient also depends on $2X\mathcal L_{XX}$. The ratio $Q=r/(\mathcal L_X+3F)$ is used whenever its denominator is positive. The coefficient $c_s^2$ describes the $k$-essence kinetic term; the effects of NMDC-modified propagation are derived in Sec.~\ref{subsec:intrinsic}.

The Friedmann equation can be written as
\begin{equation}
3H^2 = \frac{1}{M_p^2}\left(2X\mathcal{L}_X - \mathcal{L} + 9FX + \rho_r\right),
\label{eq:Friedmann}
\end{equation}
and the radiation energy density satisfies
\begin{equation}
\dot{\rho}_r + 4H\rho_r = \Gamma \dot{\phi}^2.
\label{eq:radiation}
\end{equation}

In the quasistationary radiation regime, $|\dot\rho_r|\ll4H\rho_r$, the radiation equation yields
\begin{equation}
\rho_r\simeq\frac{\Gamma\dot\phi^2}{4H}=\frac32rX.
\label{eq:rhorSR}
\end{equation}
Let $\Gamma=\Gamma_0$, so $\beta\equiv\dot\Gamma/(H\Gamma)=0$. The Friedmann relation determines the homogeneous background, whereas the local fluctuation coefficients are derived from the field action in Sec.~\ref{subsec:intrinsic}.

\subsection{Slow-roll parameters}

With no potential present, the slow evolution is driven by changes in $K(\phi)$. We define
\begin{align}
\epsilon &\equiv -\frac{\dot{H}}{H^2}, \label{eq:epsilon} \\
\eta &\equiv \frac{\dot{K}}{H K} = \frac{K_\phi \dot{\phi}}{H K}, \label{eq:eta} \\
\beta &\equiv \frac{\dot{\Gamma}}{H\Gamma} = \frac{\Gamma_\phi \dot{\phi}}{H\Gamma}. \label{eq:beta}
\end{align}
A sustained slow-roll inflationary phase requires $\epsilon \ll 1$, $|\eta| \ll 1$, and $|\beta| \ll 1$. Together with positive kinetic and gradient coefficients, these inequalities define the regime of the following perturbation analysis.

\subsection{Kinetic attractor and slow evolution}

Based on the kinetic-attractor analysis of Ref.~\cite{Zhang:2026}, we organize the slow evolution around the instantaneous quasi-de Sitter fixed point. To leading order, $\rho+p\simeq0$ yields
\begin{equation}
X_f = \mu^4\,\frac{\overline{K} - 3F - r}{2},
\label{eq:Xf}
\end{equation}
which requires $\overline{K} > 3F + r$. The associated inflaton velocity, where $\sigma=\operatorname{sgn}(\dot\phi)=\pm1$, is
\begin{equation}
\dot{\phi}_f = \sigma \mu^2 \sqrt{\overline{K} - 3F - r},
\end{equation}
and the Hubble scale is given by
\begin{equation}
H_f^2 = \frac{1}{3M_p^2}\, \frac{X_f}{2}\,(9F + \overline{K}).
\end{equation}
The slow-roll trajectory is $X_{\mathrm{sr}}=X_f+\delta X$, where $|\delta X|/X_f\ll1$. We evaluate the kinetic and relaxation coefficients along this evolving trajectory. To leading order, the number of $e$-folds with the convention $N_{,\phi}=H/\dot\phi$ is expressed as
\begin{equation}
N \simeq \frac{\sigma}{2\sqrt{3}M_p} \int \sqrt{9F + \overline{K}}\; d\phi.
\end{equation}

\subsection{Background consistency}
Let $D=\mathcal L_X+3F$, $\Delta=\overline K-3F-r>0$, and $X_{\rm sr}=X_f(1+\delta)$ with $|\delta|\ll1$. The fixed-background inertial and gradient coefficients derived below satisfy
\begin{equation}
C=(2+3\delta)\Delta-r,\qquad
G=\delta\Delta-r-2\epsilon F.
\label{eq:CGfixedtest}
\end{equation}
These coefficients are evaluated along the evolving trajectory, not exactly at $X_f$. On the branch with $\epsilon\ge0$, $F\ge0$ and $G>0$, Eq.~\eqref{eq:CGfixedtest} implies $r<\delta\Delta$, which in turn yields
\begin{equation}
\begin{gathered}
0\le\frac rC<\frac{\delta}{2(1+\delta)}\ll1,\\
\Xi\equiv\frac{CH}{3HC+\Gamma_0}
=\frac{1}{3(1+r/C)}\simeq\frac13.
\end{gathered}
\label{eq:chi_branch}
\end{equation}
Accordingly, the stable near-attractor branch requires finite-inertia propagation. The retarded formulation developed below applies directly to this branch, while Appendix~\ref{app:ODreference} separately discusses the overdamped reference response.

\section{Analysis of primordial non-Gaussianity in warm $k$-inflation with NMDC}
\label{sec:nonGaussianity}

The nonlinear $\delta N$ term and the intrinsic thermal bispectrum are calculated, and then the gravitational constraints and local cubic interactions are derived in the ADM formalism.

\subsection{General definitions and theoretical preliminaries}

$\zeta\simeq(H/\dot\phi)\delta\phi$ is used in the single-clock adiabatic approximation, and this sign convention is kept throughout. The linear uniform-inflaton spatial curvature utilized in Sec.~\ref{subsec:metric}, $h_{ij}=a^2e^{2\zeta_u}\delta_{ij}$, satisfies $\zeta_u=-\zeta$. Accordingly, the minus sign in Eq.~\eqref{eq:bispec} is consistent with the usual positive-sign bispectrum convention for $\zeta_u$.

The power spectrum of the curvature perturbation, represented as $P_{\zeta}(k)$, is defined using the two-point correlation function in Fourier space:
\begin{equation}
\langle \zeta_{\mathbf{k}_1} \zeta_{\mathbf{k}_2} \rangle = (2\pi)^3 \delta^3(\mathbf{k}_1 + \mathbf{k}_2)\, P_{\zeta}(k_1),
\label{eq:twopoint}
\end{equation}
where $\zeta_{\mathbf{k}}$ denotes the Fourier mode of $\zeta$. The dimensionless power spectrum is expressed as $\mathcal{P}_{\zeta}(k) = \frac{k^3}{2\pi^2} P_{\zeta}(k)$.

The three-point correlation function provides the leading statistic of primordial non-Gaussianity; its Fourier transform is the bispectrum $B_{\zeta}(k_1,k_2,k_3)$:
\begin{equation}
\langle \zeta_{\mathbf{k}_1} \zeta_{\mathbf{k}_2} \zeta_{\mathbf{k}_3} \rangle = (2\pi)^3 \delta^3(\mathbf{k}_1 + \mathbf{k}_2 + \mathbf{k}_3)\, B_{\zeta}(k_1,k_2,k_3).
\label{eq:threepoint}
\end{equation}
For an almost Gaussian distribution, the nonlinearity parameter $f_{\mathrm{NL}}$ is defined as
\begin{equation}
\begin{aligned}
B_{\zeta}(k_1,k_2,k_3)=-\frac65 f_{\rm NL}(k_1,k_2,k_3)\\
\times\bigl[P_{\zeta}(k_1)P_{\zeta}(k_2)+\mathrm{cyc.}\bigr],
\end{aligned}
\label{eq:bispec}
\end{equation}
where the cyclic permutations run over the three wavenumbers, and the sign convention is specified by Eq.~\eqref{eq:bispec}. For $k_1=k_2=k_3=k$, this expression becomes $B_{\zeta}(k)=-\frac{18}{5}f_{\mathrm{NL}}P_{\zeta}(k)^2$. Planck bounds on various bispectrum templates~\cite{Planck:2020,Jung:2025} therefore make non-Gaussianity a valuable discriminant among inflationary models.

With a common scalar power spectrum and curvature convention, the tree-level non-Gaussianity can be organized into intrinsic, nonlinear $\delta N$, and metric--radiation contributions,
\begin{equation}
f_{\mathrm{NL}}^{\mathrm{total}}=f_{\mathrm{NL}}^{\mathrm{int}}+f_{\mathrm{NL}}^{\delta N}+f_{\mathrm{NL}}^{\mathrm{metric/rad}}.
\label{eq:total}
\end{equation}
The intrinsic and $\delta N$ terms are evaluated below. In Section~\ref{subsec:metric}, we derive the constraints and local vertices entering the metric--radiation term; to obtain the complete bispectrum, we also need to account for the coupled dissipative transfer dynamics.

\subsection{$\delta N$ contribution}
\label{subsec:dN}

Within the single-clock approximation, the $\delta N$ expansion is
$\zeta=N_{,\phi}\delta\phi+N_{,\phi\phi}(\delta\phi)^2/2+\cdots$~\cite{Lyth:2005a,Lyth:2005b}. For Gaussian $\delta\phi$, our bispectrum convention yields
\begin{equation}
f_{\mathrm{NL}}^{\delta N}=-\frac56\frac{N_{,\phi\phi}}{N_{,\phi}^2}.
\label{eq:dN_basic}
\end{equation}
The leading kinetic fixed-point relations imply that
\begin{equation}
\begin{aligned}
N_{,\phi}&=\frac{H}{\dot\phi}
=\frac{\sigma}{2\sqrt3M_p}\sqrt{9F+\overline K},\\
N_{,\phi\phi}&=\frac{\sigma}{4\sqrt3M_p}
\frac{9F_{,\phi}+\overline K_{,\phi}}{\sqrt{9F+\overline K}}.
\end{aligned}
\label{eq:Nphi}
\end{equation}
The derivative $F_{,\phi}=\dot F/\dot\phi=-2\epsilon F H/\dot\phi$ is taken along the adiabatic background, and $\overline K_{,\phi}=\eta\overline K H/\dot\phi$. It follows that $N_{,\phi\phi}/N_{,\phi}^2=(\overline K\eta-18F\epsilon)/[2(9F+\overline K)]$, and hence
\begin{equation}
\boxed{
f_{\mathrm{NL}}^{\delta N}
=\frac5{12}\frac{K\eta+18F\epsilon}{9F+\overline K}.
}
\label{eq:fNL_dN}
\end{equation}

Equation~\eqref{eq:fNL_dN} presents the nonlinear adiabatic-conversion result. For $F\ge0$ and $\overline K>0$, the weights of $\epsilon$ and $\eta$ are bounded, keeping this contribution at first order in slow roll. Although NMDC modifies their relative weights, it does not produce an inverse-sound-speed enhancement in this term. The intrinsic interactions discussed below therefore constitute a separate source of larger, momentum-dependent signals.

\subsection{Intrinsic thermal bispectrum}
\label{subsec:intrinsic}

Hereafter, FI refers to the finite-inertia (retarded) result, FO to the local freeze-out approximation, and OD to the overdamped reference response in Appendix~\ref{app:ODreference}.

\subsubsection{Cubic action and nonlinear source}
We expand $\Phi=\phi_0(t)+u(\mathbf x,t)$ on a fixed FLRW geometry and define $v=\dot\phi_0$ and $X_0=v^2/2$. The exact local Lagrangian and current equation are presented in Appendix~\ref{app:source}. The following quadratic coefficients are introduced
\begin{align}
C&\equiv\mathcal L_X+2X_0\mathcal L_{XX}+3F
=K+\frac{6X_0}{\mu^4}+3F,\label{eq:Ckinnew}\\
G&\equiv\mathcal L_X+3F+\frac{2\dot H}{M^2},\label{eq:Ggradnew}\\
\mathcal B_\phi&\equiv K_\phi v,\qquad
m_{\rm eff}^2\equiv 3H\mathcal B_\phi+\dot{\mathcal B}_\phi-K_{\phi\phi}X_0,\label{eq:massnew}
\end{align}
and the quadratic action is written as
\begin{equation}
\begin{aligned}
S_u^{(2)}=\int dt\,d^3x\,a^3\bigg[&
\frac C2\dot u^2-\frac G{2a^2}(\nabla u)^2\\
&+K_\phi v\,u\dot u+\frac{K_{\phi\phi}X_0}{2}u^2\bigg].
\end{aligned}
\label{eq:fixed_s2}
\end{equation}
Variation leads to the linear stochastic equation
\begin{equation}
C\ddot u_1+\gamma_{\rm lin}\dot u_1
-\frac G{a^2}\nabla^2u_1+m_{\rm eff}^2u_1=\xi,
\label{eq:correct_linear}
\end{equation}
where
\begin{equation}
\gamma_{\rm lin}=3HC+\dot C+\Gamma_0.
\label{eq:correct_gamma}
\end{equation}
This expression differs from the background friction $3H(D+r)$ by a term proportional to $2X_0\mathcal L_{XX}$ and by background derivatives. In general, the noncanonical contribution is not suppressed by slow roll.

The complete \emph{fixed-metric pure-field} cubic action is given by
\begin{equation}
\begin{aligned}
S_u^{(3)}=\int\!dt\,d^3x\,a^3\bigg\{&
\frac{v}{\mu^4}\dot u\left[\dot u^2
-\frac{(\nabla u)^2}{a^2}\right]\\
&+\frac{K_\phi}{2}u\left[\dot u^2
-\frac{(\nabla u)^2}{a^2}\right]\\
&+\frac{K_{\phi\phi}v}{2}u^2\dot u
+\frac{K_{\phi\phi\phi}X_0}{6}u^3\bigg\}.
\end{aligned}
\label{eq:correct_cubic}
\end{equation}
Specifically, the two $v/\mu^4$ vertices encompass the unsuppressed temporal and spatial-gradient kinetic nonlinearities. The NMDC term in Eq.~\eqref{eq:fixedL} is quadratic in $\Phi$ and does not have \emph{direct} pure-field cubic vertices on fixed FLRW. It still contributes through $C$, $G$, the background solution, and the stochastic mode functions. Its cubic couplings involving metric perturbations are discussed separately in Sec.~\ref{subsec:metric}.

The second-order field perturbation obeys
\begin{equation}
\begin{aligned}
\mathscr L u_2&=S_2[u_1],\\
S_2&=-(\partial_t+3H)J_2
+\frac{\nabla\cdot\mathbf V_2}{a^2}+P_{\phi,2}.
\end{aligned}
\label{eq:correct_s2source}
\end{equation}
Here, $\mathscr L$ denotes the linear operator of Eq.~\eqref{eq:correct_linear}, and $u=u_1+u_2$ has no factor $1/2$ in $u_2$. The explicit quadratic currents $J_2,\mathbf V_2,P_{\phi,2}$, derived by varying Eq.~\eqref{eq:correct_cubic}, are given in Appendix~\ref{app:source}. This expression preserves the time-current derivative and the complete spatial-current divergence.

\subsubsection{Finite-inertia bispectrum}
Let $R_k(t,s)$ represent the retarded inverse of $\mathscr L_k$, and let $\mathscr C_k(t,s)$ represent the unequal-time Gaussian field covariance. We impose $u_2(t_0)=\dot u_2(t_0)=0$ and assume an initial Gaussian field--velocity state that is independent of later additive noise, with noise strength $2\Gamma_0Ta^{-3}\delta(t-s)$. Contracting the quadratic source with two external fields yields $\mathscr S_{ij|\ell}(t,s)$, whose explicit evaluation is provided in Appendix~\ref{app:source}.

Given vanishing initial second-order data, the intrinsic fixed-background tree-level field bispectrum is therefore represented as
\begin{equation}
\boxed{
\begin{aligned}
B_u(k_1,k_2,k_3;t)&=\sum_{\mathrm{cyc}(ij|\ell)}\int_{t_0}^{t}\!ds\\
&\quad\times R_{k_\ell}(t,s)\,\mathscr S_{ij|\ell}(t,s).
\end{aligned}
}
\label{eq:formal_B}
\end{equation}
The momentum-conserving delta function $(2\pi)^3\delta^3(\mathbf k_1+\mathbf k_2+\mathbf k_3)$ has been eliminated from $B_u$. Equation~\eqref{eq:formal_B} preserves the full momentum triangle and the unequal-time correlations.

Adopting the same single-clock conversion and bispectrum sign convention as above, we have
\begin{equation}
B_\zeta^{\rm int}=\left(\frac H v\right)^3 B_u,
\qquad
f_{\rm NL}^{\rm int}
=-\frac56\frac{B_\zeta^{\rm int}}
{P_\zeta(k_1)P_\zeta(k_2)+\mathrm{cyc.}}.
\label{eq:fnl_formal_new}
\end{equation}
For equilateral momenta, all three cyclic terms are identical. The finite-inertia result is then
\begin{equation}
\boxed{
f_{\rm NL,eq}^{\rm int,FI}(k;t)
=-\frac{5v(t)}{6H(t)P_u(k,t)^2}\,\mathcal I_k(t),
}
\label{eq:FI_equilateral}
\end{equation}
where the complete response weight is
\begin{equation}
\begin{aligned}
\mathcal I_k(t)={}&R_k(t,t_0)\mathscr J_{\rm eq}(t,t_0)\\
&+\int_{t_0}^t ds\,\bigg[(\partial_s R_k-3HR_k)
\mathscr J_{\rm eq}\\
&\qquad+R_k\left(\mathscr P_{\rm eq}
+\frac{i\mathbf k_3\cdot\bm{\mathscr V}_{\rm eq}}{a^2}\right)\bigg].
\end{aligned}
\label{eq:FI_weight}
\end{equation}
Here, $P_u(k,t)=\mathscr C_k(t,t)$, and the pairwise currents on the right are represented in Eqs.~\eqref{eq:WickJ}--\eqref{eq:WickV}, with $k_1=k_2=k_3=k$. The response weight $\mathcal I_k$ accounts for both the unequal-time propagation and the redshift of physical momenta. It contains the two Wick pairings associated with one source leg; the three equal cyclic terms cancel out the factor of three in the equilateral power-spectrum denominator.

The integrated-by-parts representation in Eq.~\eqref{eq:finiteIBP} provides an alternative expression for the same result. Its lower endpoint preserves the initial Gaussian covariance, whereas the finite-inertia condition $R_k(t,t)=0$ eliminates the upper endpoint.

\subsubsection{Joint power spectrum and bispectrum}
For $\mathbf y_k=(u_k,\dot u_k)^T$, let $\boldsymbol{\Sigma}_k$ be the Gaussian spectral covariance matrix whose entries are $\Sigma_{k,ab}$, with the momentum-conserving delta function removed. Let $\mathsf U_k(t,s)$ be the homogeneous evolution matrix. The noise-generated contribution is obtained by setting $\boldsymbol{\Sigma}_k(t_0)=0$, while any nonzero thermal initial covariance evolves as $\mathsf U_k(t,t_0)\boldsymbol{\Sigma}_k(t_0)\mathsf U_k^T(t,t_0)$. Both contributions are included in the covariance equation in Appendix~\ref{app:validation}. Its solution yields $P_u(k,t)=\Sigma_{k,11}(t)$, with the same state and mode normalization as $B_u$. For the purely thermal sector, the single-clock conversion leads to
\begin{equation}
\begin{aligned}
\mathcal P_\zeta^{\rm th}(k)&=\left(\frac Hv\right)^2\frac{k^3}{2\pi^2}\Sigma_{k,11},\\
f_{\rm NL}^{\rm int,th}&=-\frac{5v}{6H}
\frac{B_u(k_1,k_2,k_3)}{\sum_{\rm cyc}\Sigma_{k_i,11}\Sigma_{k_j,11}}.
\end{aligned}
\label{eq:joint_fNL}
\end{equation}
The covariance evolution and unequal-time reconstruction are presented in Appendix~\ref{app:validation}. Equation~\eqref{eq:joint_fNL} adopts a common stochastic normalization for both correlators; ``th'' signifies thermal-spectrum normalization. When comparing against a total scalar spectrum, the total power in Eq.~\eqref{eq:fnl_formal_new} should be used.

\subsubsection{Local amplitude--shape reduction}
For modes with comparable momenta, we apply a local common-freeze-out closure with $\gamma_d=3HC+\Gamma_0$ and
\begin{equation}
\lambda_k=\frac{Gk^2/a^2+m_{\rm eff}^2}{\gamma_d},\qquad
\lambda_{k_*}(t_F)\simeq H.
\label{eq:lambdaFO}
\end{equation}
Here, $k$ denotes a comoving momentum, and $k_*$ represents the reference comoving momentum. For positive $G$ and $\gamma_d H>m_{\rm eff}^2$, the physical freeze-out scale and, when $k_F$ exceeds $H$, its horizon-crossing window are
\begin{equation}
k_F^2\simeq\frac{\gamma_d H-m_{\rm eff}^2}{G},\qquad
L_F\equiv\ln\frac{k_F}{H}>0.
\label{eq:kFOnew}
\end{equation}
Based on the common freeze-out-time treatment of thermal fluctuations~\cite{Berera:2000,Gupta:2002}, this study takes $\Delta t_F=L_F/H$ and evaluates the slowly varying vertex coefficients and their first background-time derivatives at one reference time while setting $H$ constant in the response exponents. Defining $P_i(t)=P_u(k_i,t)$, we use the local covariance-response closure $\mathscr C_{k_i}(t,s)\simeq P_i(t)e^{-H(t-s)}$ and $R_{k_\ell}(t,s)\simeq\gamma_d^{-1}e^{-H(t-s)}$ for near-equal-scale modes. We retain $a(s)^{-2}=a(t)^{-2}e^{2H(t-s)}$ in the spatial vertices. Appendix~\ref{app:validation} compares this local closure with the finite-inertia calculation.

With the finite-window response factors $L_0=(1-e^{-3L_F})/3$ and $L_g=1-e^{-L_F}$, we define the power-weighted geometric shapes ($\mathbf k_i+\mathbf k_j+\mathbf k_\ell=0$) as
\begin{equation}
\begin{aligned}
\mathfrak S_1&=\frac{\sum_{\rm cyc}P_iP_j\,\mathbf k_i\!\cdot\!\mathbf k_j}
{k_*^2\sum_{\rm cyc}P_iP_j},\\
\mathfrak S_2&=\frac{\sum_{\rm cyc}P_iP_j\,k_\ell^2}
{k_*^2\sum_{\rm cyc}P_iP_j}.
\end{aligned}
\label{eq:FOshapes}
\end{equation}
Substituting the complete current source, Eqs.~\eqref{eq:WickJ}--\eqref{eq:WickV}, into the retarded integral yields the compact nonlinearity parameter
\begin{equation}
\boxed{
\begin{aligned}
f_{\rm NL}^{\rm int,FO}\simeq{}&L_0\,\mathcal A_{\rm FO}\\
&+L_g\left(\mathcal C_{1,\rm FO}\mathfrak S_1
+\mathcal C_{2,\rm FO}\mathfrak S_2\right).
\end{aligned}
}
\label{eq:FOmaster}
\end{equation}
The amplitude $\mathcal A_{\rm FO}$ and shape coefficients $\mathcal C_{1,\rm FO},\mathcal C_{2,\rm FO}$ are presented in Appendix~\ref{app:compactFO}.
The vertices utilize the reference background at $t_F$, whereas $k_*^2/a(t)^2$ and the curvature-conversion factor $v(t)/H(t)$ are defined on the common terminal surface. For an equilateral triangle $\mathfrak S_1=-1/2$ and $\mathfrak S_2=1$, yielding
\begin{equation}
f_{\rm NL,eq}^{\rm int,FO}\simeq
L_0\mathcal A_{\rm FO}
+L_g\left(-\frac12\mathcal C_{1,\rm FO}
+\mathcal C_{2,\rm FO}\right).
\label{eq:FOeq}
\end{equation}
In the further short-window expansion $L_F\ll1$, we have $L_0=L_F+O(L_F^2)$ and $L_g=L_F+O(L_F^2)$, and this recovers the familiar
$L_F[\mathcal A+\mathcal C_1\mathfrak S_1+\mathcal C_2\mathfrak S_2]$ structure of the original freeze-out treatment. The two different finite-window weights in Eq.~\eqref{eq:FOmaster} capture the leading impact of spatial redshift. The derivation, including the intermediate contracted current, is presented in Appendix~\ref{app:compactFO}.

The main intrinsic results are the full-triangle retarded bispectrum, Eq.~\eqref{eq:formal_B}, and the corresponding equilateral nonlinearity parameter, Eq.~\eqref{eq:FI_equilateral}. They incorporate finite inertia, unequal-time propagation, and all fixed-FLRW cubic field vertices, with the two-point function computed from the same state. Equation~\eqref{eq:FOmaster} offers a complementary analytic separation: $\mathcal A_{\rm FO}$ multiplies the momentum-independent term, whereas $\mathcal C_{1,\rm FO}$ and $\mathcal C_{2,\rm FO}$ weight the two geometric contractions. The same-background test described in Appendix~\ref{app:validation} confirms that the finite-inertia expression provides the quantitative result on the tested near-attractor branch.

\subsection{Metric constraints and derivative-coupling interactions}
\label{subsec:metric}

The derivative coupling alters the lapse and shift constraints even before the introduction of stochastic matter dynamics. We adopt the ADM decomposition
\begin{equation}
 ds^2=-N^2dt^2+h_{ij}(dx^i+N^idt)(dx^j+N^jdt),
\end{equation}
with $K_{ij}=(\dot h_{ij}-D_iN_j-D_jN_i)/(2N)$. In the uniform-inflaton gauge, $\phi=\phi_0(t)$, define $X_N=\dot\phi_0^2/(2N^2)$. The Gauss--Codazzi relation then yields
\begin{equation}
G^{\mu\nu}\partial_\mu\phi\partial_\nu\phi
=X_N\left[{}^{(3)}R+K^2-K_{ij}K^{ij}\right].
\end{equation}
It follows that the exact gravitational plus NMDC action, up to the Einstein--Hilbert boundary term, is
\begin{equation}
\begin{aligned}
S_{g+\mathrm{NMDC}}={}&\frac12\int\!dt\,d^3x\,N\sqrt h\\
&\times\Biggl[\left(M_p^2+\frac{X_N}{M^2}\right){}^{(3)}R\\
&\quad+\left(M_p^2-\frac{X_N}{M^2}\right)
(K_{ij}K^{ij}-K^2)\Biggr].
\end{aligned}
\label{eq:ADMexact}
\end{equation}
The scalar and radiation actions must be included before the scalar constraints are imposed. With $Z=X/(M_p^2M^2)$, the tensor coefficients are
\begin{equation}
\mathcal G_T=M_p^2(1-Z),\qquad
\mathcal F_T=M_p^2(1+Z).
\end{equation}
Thus, $0\le Z<1$ is required for tensor stability.

To assess the magnitude of NMDC mixing at the constraint level, we consider the scalar-only part of the lapse constraint while setting the radiation momentum perturbation to zero in Eq.~\eqref{eq:lapseFluid}. When $\Theta\ne0$, this gives
\begin{equation}
\alpha_{\rm sc}=\frac{1-Z}{1-3Z}\frac{\dot\zeta_u}{H}
=\left(1+\frac{2Z}{1-3Z}\right)\frac{\dot\zeta_u}{H}.
\label{eq:lapseEstimate}
\end{equation}
For $|Z|\ll1$ and $|1-3Z|\sim1$, the NMDC correction to the linear constraint is $2Z/(1-3Z)=2Z+O(Z^2)$, while the tensor coefficients receive $O(Z)$ corrections. The Einstein-gravity reference utilizes the same noncanonical kinetic Lagrangian. Given that $Z/F=X/(M_p^2H^2)$, metric mixing and background friction characterize distinct aspects of the derivative coupling. Consequently, the constraint correction may be small even when NMDC substantially affects the background and intrinsic response. Derivative coupling also introduces local cubic metric interactions~\cite{DeFelice:2011,Kobayashi:2011}, which are derived in the following.

The scalar-only quadratic coefficients $\mathcal G_S,\mathcal F_S$ and the Horndeski combinations $\Theta,\Sigma_{\rm H}$ are given in Appendix~\ref{app:coupled}.
They characterize the constrained scalar action and are distinct from $C,G$ in the fixed-background Langevin equation. The unitary-gauge constraint for $N=1+\alpha$, $N_i=\partial_i\chi$ and $h_{ij}=a^2e^{2\zeta_u}\delta_{ij}$ is given by
\begin{equation}
\Theta\alpha=\mathcal G_T\dot\zeta_u
+\frac12\dot\sigma P_{r,Y}\,\pi
\label{eq:lapseFluid}
\end{equation}
for the minimal radiation realization $P_r(Y)$ with $\sigma=\sigma_0+\pi$ and $Y=-g^{\mu\nu}\partial_\mu\sigma\partial_\nu\sigma/2$.
Thus, the radiation momentum perturbation appears explicitly in the lapse constraint.

For the minimal barotropic radiation model $P_r(Y)=\lambda Y^2$ with $\lambda>0$, the two propagating high-frequency scalar modes satisfy
\begin{equation}
 c_r^2=\frac13,\qquad
 c_{\zeta}^2=\frac{\displaystyle \mathcal F_S-
 \frac{\rho_r+p_r}{2}\left(\frac{\mathcal G_T}{\Theta}\right)^2}
 {\mathcal G_S}.
\label{eq:coupledcs}
\end{equation}
In this closure, the corresponding necessary principal-part stability conditions are
\begin{equation}
\mathcal G_S>0,\quad
\mathcal F_S>\frac{\rho_r+p_r}{2}
\left(\frac{\mathcal G_T}{\Theta}\right)^2,
\quad\rho_r+p_r>0,
\label{eq:conditionalstability}
\end{equation}
with $0\le Z<1$ and $\Theta\ne0$. These high-frequency conditions are obtained from the constrained quadratic action derived in Appendix~\ref{app:coupled}. Meanwhile, the dissipative growth and relaxation rely on the transfer operator.

The local gravity--inflaton cubic action can also be expanded directly. The off-shell terms in unitary gauge, including the kinetic and extrinsic-curvature contributions, are presented in Appendix~\ref{app:cubic}. The spatial-curvature part involves
\begin{equation}
\boxed{
\begin{aligned}
S_{R_3}^{(3)}={}&M_p^2\!\int\!dt\,d^3x\,a\\
&\times\bigl[(1+Z)\zeta_u(\partial\zeta_u)^2\\
 &+(1-Z)\alpha(\partial\zeta_u)^2\\
 &+2(1-Z)\zeta_u\partial_i\alpha\partial_i\zeta_u\\
 &+4Z\alpha\partial_i\alpha\partial_i\zeta_u\bigr].
\end{aligned}
}
\label{eq:metriccubicR3}
\end{equation}
The intrinsic calculation reveals the thermal field self-interactions, whereas the ADM analysis determines the direct NMDC corrections to the gravitational constraints and cubic vertices~\cite{Maldacena:2003,Kobayashi:2011,DeFelice:2011,Kase:2014}. The lapse mixing in Eq.~\eqref{eq:lapseEstimate} and the $4Z\alpha\partial_i\alpha\partial_i\zeta_u$ vertex in Eq.~\eqref{eq:metriccubicR3} distinguish these gravitational effects from the modified field propagation. Along with Appendix~\ref{app:cubic}, they provide the local interactions required for a coupled metric--radiation bispectrum. Determining its amplitude additionally requires the constrained radiation dynamics and dissipative transfer kernels.

\section{Physical interpretation and implications}
\label{sec:results}

The three components of the calculation serve distinct physical purposes. Equation~\eqref{eq:fNL_dN} describes the local nonlinear adiabatic conversion and is slow-roll suppressed. Equations~\eqref{eq:formal_B} and \eqref{eq:FI_equilateral} contain the main intrinsic-bispectrum results: the same cubic source determines both the full triangle dependence and its equilateral amplitude. The ADM results, Eqs.~\eqref{eq:lapseEstimate} and \eqref{eq:metriccubicR3}, characterize the associated constraint mixing and local gravitational interactions.

The $X^2/\mu^4$ interaction generates both $v\dot u^3/\mu^4$ and $-v\dot u(\nabla u)^2/(\mu^4a^2)$. These temporal and gradient vertices make different contributions to the bispectrum amplitude and momentum dependence. The derivatives $K_\phi,K_{\phi\phi},K_{\phi\phi\phi}$ introduce additional field-dependent interactions. For the fixed-background propagation speed $c_u^2=G/C$, the identity $4X/(\mu^4G)=c_u^{-2}-1-2\epsilon F/G$ links the kinetic interaction strength to the sound speed and NMDC correction, and its contribution to the bispectrum is weighted by the retarded propagation. On a fixed FLRW geometry, NMDC does not have direct cubic pure-field vertices; instead, it affects the intrinsic bispectrum through the background and the propagation coefficients $C,G,\gamma_{\rm lin}$. Its direct local gravitational vertices are shown separately by the ADM expansion.

In Eq.~\eqref{eq:FOmaster}, the response weights $L_0$ and $L_g$ capture the distinct redshift dependence of temporal and spatial-gradient interactions. The coefficients $\mathcal C_{1,\rm FO}$ and $\mathcal C_{2,\rm FO}$ determine their coupling strengths, while $\mathfrak S_1$ and $\mathfrak S_2$ encode the momentum geometry. In the equilateral limit, these shapes become $-1/2$ and $1$, yielding the single combination in Eq.~\eqref{eq:FOeq}. This compact formula therefore illustrates how the interaction strengths and triangle geometry are combined in the local approximation. On the stable near-attractor branch, $\Xi\simeq1/3$ makes finite inertia relevant. In Appendix~\ref{app:validation}, the same-background comparison indicates that the local closure misses both the magnitude and sign of the finite-inertia equilateral result on the reference trajectory. The unequal-time retarded expression provides the quantitative intrinsic prediction in this case. Appendix~\ref{app:ODreference} additionally presents an analytic series for a distinct, constant-coefficient overdamped reference problem.

Owing to the common stochastic normalization in Eq.~\eqref{eq:joint_fNL}, the joint calculations of the power spectrum and bispectrum are reproducible for a specified Gaussian initial state. An observational comparison requires the total curvature bispectrum and scalar power spectrum, followed by projection of the bispectrum onto CMB templates~\cite{Planck:2020}. These results complement the leading-order scalar-spectrum analysis in Ref.~\cite{Zhang:2026}, resolving the nonlinear interactions and response history that generate the intrinsic three-point function.

\section{Conclusions}
\label{conclusion}

In this study, the intrinsic thermal bispectrum of potential-free warm $k$-inflation with NMDC has been derived from the complete fixed-FLRW cubic inflaton action. Our main result is the finite-inertia retarded expression in Eq.~\eqref{eq:formal_B}, and the corresponding equilateral nonlinearity parameter is given in Eq.~\eqref{eq:FI_equilateral}. This formulation incorporates temporal, gradient, and field-dependent kinetic vertices, the initial Gaussian covariance, as well as unequal-time stochastic propagation. The two- and three-point functions utilize the same response and statistical normalization.

The compact amplitude--shape relation in Eq.~\eqref{eq:FOmaster} and its equilateral limit in Eq.~\eqref{eq:FOeq} isolate the cubic couplings from the triangle geometry and offer a transparent local analytic estimate. Meanwhile, the finite-inertia expression in Eq.~\eqref{eq:FI_equilateral} preserves the unequal-time propagation needed for the quantitative amplitude on the tested near-attractor branch. Moreover, the nonlinear $\delta N$ term is slow-roll suppressed, while the ADM analysis identifies the corrections to the derivative-coupling constraint, along with local gravitational vertices. These findings extend our earlier analysis on warm $k$-inflation non-Gaussianity to the potential-free NMDC model, linking its nonlinear kinetic interactions and modified propagation to the intrinsic thermal bispectrum.

\section*{Acknowledgements}
This work was supported by the Shandong Provincial Natural Science Foundation (Grant Nos.~ZR2021MA037 and ZR2022JQ04) and the National Natural Science Foundation of China (Grant Nos.~12575134 and 11605100).
\section*{Statements and Declarations}
\noindent\textit{Funding.} This research was funded by the grants acknowledged above.

\noindent\textit{Data availability.} The analytic derivations, numerical inputs, and reference results are included in the main text and appendices. The verification scripts and machine-readable outputs are available from the corresponding author upon reasonable request.

\noindent\textit{Competing interests.} The authors declare no competing interests.

\appendix
\numberwithin{equation}{section}

\section{Cubic source and unequal-time contractions}
\label{app:source}

This appendix supplies the field equation, quadratic currents, and Gaussian contractions used in Sec.~\ref{subsec:intrinsic}.

Given that $G^{00}=3H^2$ and $G^{ij}=-(3H^2+2\dot H)\delta^{ij}/a^2$, the exact local field Lagrangian on this prescribed geometry is given by
\begin{equation}
\begin{aligned}
\mathcal L_{\rm fixed}={}&K(\Phi)X+\frac{X^2}{\mu^4}
+\frac{3F}{2}\dot\Phi^2\\
&-\frac{3F+2\dot H/M^2}{2a^2}(\nabla\Phi)^2,\\
X={}&\frac12\big[\dot\Phi^2-a^{-2}(\nabla\Phi)^2\big].
\end{aligned}
\label{eq:fixedL}
\end{equation}
We characterize thermal dissipation using a constant damping coefficient $\Gamma_0$ and additive Gaussian noise $\xi$. Define $\mathscr D(\Phi,X)=\mathcal L_X+3F$ and
$\mathscr G(\Phi,X)=\mathscr D+2\dot H/M^2$. Then, the exact local field equation is
\begin{equation}
\begin{aligned}
(\partial_t+3H)&[\mathscr D\dot\Phi]
-\frac1{a^2}\partial_i(\mathscr G\partial_i\Phi)\\
&-\mathcal L_{\Phi}+\Gamma_0\dot\Phi=\xi.
\end{aligned}
\label{eq:correct_current}
\end{equation}
Here, $F=F(t)$ is set by the homogeneous geometry.

Using $u=u_1+u_2$ with no factor $1/2$ in $u_2$, the quadratic time current, spatial current, and explicit field derivative are defined by
\begin{align}
J_2={}&\frac{K_{\phi\phi}v}{2}u_1^2+K_\phi u_1\dot u_1
+\frac{3v}{\mu^4}\dot u_1^2
-\frac{v}{\mu^4a^2}(\nabla u_1)^2,\label{eq:J2new}\\
\mathbf V_2={}&\left(K_\phi u_1+\frac{2v}{\mu^4}\dot u_1\right)\nabla u_1,\label{eq:V2new}\\
P_{\phi,2}={}&\frac{K_{\phi\phi\phi}X_0}{2}u_1^2
+K_{\phi\phi}v u_1\dot u_1\nonumber\\
&+\frac{K_\phi}{2}\left[\dot u_1^2-\frac{(\nabla u_1)^2}{a^2}\right].\label{eq:P2new}
\end{align}

The second-order equation is shown in Eq.~\eqref{eq:correct_s2source}. Its retarded Green function solves
\begin{equation}
\mathscr L_k R_k(t,s)=\delta(t-s),\quad
R_k(t,s)=0\quad(t<s).
\label{eq:retarded_new}
\end{equation}
The second-order solution, for $u_2(t_0)=\dot u_2(t_0)=0$, is given by
\begin{equation}
u_{2,\mathbf k}(t)=\int_{t_0}^{t}ds\,R_k(t,s)S_{2,\mathbf k}(s).
\label{eq:u2retnew}
\end{equation}
For Gaussian additive noise that satisfies
\begin{equation}
\langle\xi_{\mathbf k}(t)\xi_{\mathbf k'}(s)\rangle
=(2\pi)^3\delta^3(\mathbf k+\mathbf k')
2\Gamma_0T(t)a^{-3}(t)\delta(t-s),
\label{eq:noisenew}
\end{equation}
the unequal-time covariance is defined as follows
\begin{equation}
\langle u_{1,\mathbf k}(t)u_{1,\mathbf k'}(s)\rangle
=(2\pi)^3\delta^3(\mathbf k+\mathbf k')\mathscr C_k(t,s).
\label{eq:Cdefinition}
\end{equation}
Assuming that the Gaussian initial conditions and later noise have statistical independence, the covariance involves an initial-data term and the following noise-sourced component:
\begin{equation}
\begin{aligned}
\mathscr C_k^{\rm noise}(t,s)
=2\Gamma_0\!\int_{t_0}^{\min(t,s)}\!d\tau\,
&T(\tau)a^{-3}(\tau)\\
&\times R_k(t,\tau)R_k(s,\tau).
\end{aligned}
\label{eq:Cnoise}
\end{equation}
The two-time covariance encodes the Gaussian contractions entering the bispectrum.

We now provide an explicit, momentum-dependent expression of the Wick contraction of Eq.~\eqref{eq:correct_s2source}. For an external pair $(i,j)$ at time $t$ and contracted with the source at $s$, put $C_i\equiv\mathscr C_{k_i}(t,s)$, $\dot C_i\equiv\partial_s\mathscr C_{k_i}(t,s)$ and $P_{ij}\equiv\mathbf k_i\cdot\mathbf k_j$. Then, all coefficients in the following expressions are evaluated at $s$:
\begin{align}
\mathscr J_{ij}={}&K_{\phi\phi}v C_iC_j
+K_\phi(C_i\dot C_j+\dot C_i C_j)\nonumber\\
&+\frac{6v}{\mu^4}\dot C_i\dot C_j
+\frac{2vP_{ij}}{\mu^4a^2}C_iC_j,\label{eq:WickJ}\\
\mathscr P_{ij}={}&K_{\phi\phi\phi}X_0C_iC_j
+K_{\phi\phi}v(C_i\dot C_j+\dot C_iC_j)\nonumber\\
&+K_\phi\dot C_i\dot C_j
+\frac{K_\phi P_{ij}}{a^2}C_iC_j,\label{eq:WickP}\\
\bm{\mathscr V}_{ij}={}&-iK_\phi(\mathbf k_i+\mathbf k_j)C_iC_j\nonumber\\
&-\frac{2iv}{\mu^4}(\mathbf k_j\dot C_i C_j
+\mathbf k_i C_i\dot C_j).\label{eq:WickV}
\end{align}
The factor $2$ in the last term of Eq.~\eqref{eq:WickJ} counts the two connected Wick pairings. Each pairing of the external fields with the two gradient factors yields $-P_{ij}C_iC_j$; summing the two pairings and multiplying by $-v/(\mu^4a^2)$ gives $2vP_{ij}C_iC_j/(\mu^4a^2)$. Here, $P_{ij}=\mathbf k_i\cdot\mathbf k_j$ includes no extra factor of two.

For the external triangle $\mathbf k_\ell=-\mathbf k_i-\mathbf k_j$, the contracted quadratic source is given by
\begin{equation}
\begin{aligned}
\mathscr S_{ij|\ell}(t,s)={}&
-(\partial_s+3H)\mathscr J_{ij}
+\mathscr P_{ij}\\
&+\frac{i\mathbf k_\ell\cdot\bm{\mathscr V}_{ij}}{a^2}.
\end{aligned}
\label{eq:contractedS}
\end{equation}
For the second-order (in time) Langevin operator, we have $R_k(t,t)=0$ and $\partial_tR_k(t,s)|_{t=s^+}=1/C(s)$. Therefore, the time-current derivative can be integrated by parts before the noise average is taken. This yields an exactly equivalent finite-inertia representation:
\begin{equation}
\begin{aligned}
B_u={}&\sum_{\rm cyc(ij|\ell)}\Bigg\{
R_{k_\ell}(t,t_0)\mathscr J_{ij}(t,t_0)\\
&+\int_{t_0}^t ds\,\Big[(\partial_sR_{k_\ell}-3HR_{k_\ell})\mathscr J_{ij}\\
&\quad+R_{k_\ell}\Big(\mathscr P_{ij}
+\frac{i\mathbf k_\ell\!\cdot\!\bm{\mathscr V}_{ij}}{a^2}\Big)\Big]\Bigg\}.
\end{aligned}
\label{eq:finiteIBP}
\end{equation}
The lower endpoint captures the initial Gaussian data, whereas the condition $R_k(t,t)=0$ eliminates the upper endpoint. Transferring the time derivative to the response leaves only continuous field and velocity processes, so Eq.~\eqref{eq:finiteIBP} is well defined for additive white noise. Its consistency with Eq.~\eqref{eq:formal_B} is established independently in Appendix~\ref{app:validation}.

\section{Derivation of the common-freeze-out amplitude--shape formula}
\label{app:compactFO}

We obtain Eq.~\eqref{eq:FOmaster} from the complete action-based current. In this reduction, the field-velocity contractions are interpreted as one-sided covariance derivatives inherited from the finite-inertia or regulated-noise theory, rather than the square of the formal time derivative of a first-order white-noise process. Set $\tau=t-s$ and denote $P_i=P_u(k_i,t)$. In the local near-equal-scale freeze-out closure used in Sec.~\ref{subsec:intrinsic}, we have
\begin{equation}
C_i=P_ie^{-H\tau},\quad
\dot C_i=H C_i,\quad
a(s)^{-2}=a(t)^{-2}e^{2H\tau}.
\label{eq:FOlocal}
\end{equation}
Substituting Eq.~\eqref{eq:FOlocal} into the Wick contractions in Eqs.~\eqref{eq:WickJ}--\eqref{eq:WickV} and expanding the time-current derivative decomposes the contracted source into temporal and momentum-dependent terms:
\begin{equation}
\begin{aligned}
\frac{\mathscr S_{ij|\ell}(t,s)}{P_iP_j}
={}&\mathcal T_0\,e^{-2H\tau}\\
&+\frac1{a(t)^2}
\left[\mathcal T_1(\mathbf k_i\!\cdot\!\mathbf k_j)+\mathcal T_2k_\ell^2\right],
\end{aligned}
\label{eq:FOsource}
\end{equation}
Convolving this source with the local retarded response yields the weights $L_0$ and $L_g$, as given in Eq.~\eqref{eq:FOprimitive} below. Matching the resulting expression to the shapes in Eq.~\eqref{eq:FOshapes} then determines the coefficients in Eq.~\eqref{eq:FOmaster}:
\begin{equation}
\begin{aligned}
\mathcal A_{\rm FO}&=-\frac{5v_t}{6H_t\gamma_dH_F}\mathcal T_0,\\
\mathcal C_{1,\rm FO}&=-\frac{5v_tk_*^2}{6H_t\gamma_dH_Fa(t)^2}\mathcal T_1,\\
\mathcal C_{2,\rm FO}&=-\frac{5v_tk_*^2}{6H_t\gamma_dH_Fa(t)^2}\mathcal T_2,
\end{aligned}
\label{eq:FOcoeffs}
\end{equation}
with vertex combinations
\begin{equation}
\begin{aligned}
\mathcal T_0={}&-K_{\phi\phi\phi}X_0
-K_{\phi\phi}(5Hv+\dot v)\\
&-K_\phi(9H^2+2\dot H)\\
&-\frac{6H}{\mu^4}
(5vH^2+H\dot v+2v\dot H),\\
\mathcal T_1={}&K_\phi-\frac{6Hv+2\dot v}{\mu^4},\\
\mathcal T_2={}&-K_\phi-\frac{2Hv}{\mu^4}.
\end{aligned}
\label{eq:FOvertices}
\end{equation}
Here, $H_F=H(t_F)$ sets the local response rate, $v_t=v(t)$ and $H_t=H(t)$ specify the terminal curvature conversion, and the vertex combinations adopt background values at $t_F$. In the strictly constant-background limit, $v_t/H_t=v_F/H_F$. The scale factor $a(t)$ is also terminal, as prescribed by Eq.~\eqref{eq:FOlocal}.

To derive $\mathcal T_0$ and $\mathcal T_1$, the vertex coefficients are first differentiated before they are evaluated locally: $\dot K_\phi=K_{\phi\phi}v$ and $\dot K_{\phi\phi}=K_{\phi\phi\phi}v$. Meanwhile, $\dot v$ and $\dot H$ are retained in these coefficient derivatives. Within the local closure, the term $\partial_s(a^{-2}C_iC_j)$ vanishes on its own, as covariance and physical-gradient redshifts cancel each other out; however, the derivative of its prefactor $v$ remains. The retarded integral therefore requires only
\begin{equation}
\begin{aligned}
\int_0^{\Delta t_F}\!d\tau\,\frac{e^{-3H\tau}}{\gamma_d}
&=\frac{L_0}{\gamma_d H},\\
\int_0^{\Delta t_F}\!d\tau\,\frac{e^{-H\tau}}{\gamma_d}
&=\frac{L_g}{\gamma_d H}.
\end{aligned}
\label{eq:FOprimitive}
\end{equation}
The two Gaussian pairings are already included in the coefficients of Eqs.~\eqref{eq:WickJ}--\eqref{eq:WickV}. Substituting Eq.~\eqref{eq:FOsource} into Eq.~\eqref{eq:formal_B} using $P_\zeta=(H_t/v_t)^2P_u$ on the terminal surface and normalizing by the weighted two-point sum yields Eq.~\eqref{eq:FOmaster}, with coefficients specified in Eq.~\eqref{eq:FOcoeffs}. This approximation assumes that the three legs share a common relaxation rate $H_F$ and retains the explicit cubic momentum contractions in $\mathfrak S_1$ and $\mathfrak S_2$. It applies to triangles with nearly equal scales; the general unequal-time evolution is presented in Eq.~\eqref{eq:formal_B}.

\section{Checks of the stochastic and analytic calculations}
\label{app:validation}

\subsection{Reference background and matching prescription}
We compare Eqs.~\eqref{eq:FOeq} and \eqref{eq:FI_equilateral} on one evolving 60-$e$-fold trajectory. In units $M_p=1$, we take $K(\phi)=-50+0.6\phi$, $H_0=M=10^{-5}$, $\Gamma_0=0.03H_0$, $g_*=100$, and $\mu^4/H_0^2\simeq4.186\times10^{-3}$. At $N=\ln[a/a(0)]=0$, we set $\phi=0$, $\dot\phi(0)=+\sqrt{2X(0)}$, $X/H_0^2\simeq9.932\times10^{-2}$ and $\rho_r/H_0^2=1.5r_0X/H_0^2$, with $r_0=\Gamma_0/(3H_0)$. These displayed inputs are rounded, and their unrounded values are fixed by $H(0)=H_0$ and $X(0)=1.01X_f(0)$ using Eqs.~\eqref{eq:Friedmann} and \eqref{eq:Xf}. Solving the homogeneous field and radiation equations together with the differentiated Friedmann constraint yields a trajectory with $\epsilon<1$, $T/H>1$, $C,G>0$, and positive conservative ADM--fluid principal coefficients.

At each selected $N_F$, we rescale $a(N_F)=1$ and then select the comoving equilateral mode $k=k_F(N_F)$ according to Eq.~\eqref{eq:kFOnew}. Both calculations terminate on the same horizon-crossing surface, $k/a(t)=H(t)$. We use the same terminal thermal power and $v(t)/H(t)$ for curvature conversion. The explicit FO gradient factor is $k^2/a(t)^2=k^2e^{-2(N_{\rm end}-N_F)}$, as prescribed by Eq.~\eqref{eq:FOlocal}; only the background vertices are held fixed at $N_F$. In these comparisons, the FO time window is $L_F/H_F$, differing from the exact elapsed proper time by at most $2.9\%$. By contrast, FI follows the evolving background and its additive noise.

\begin{samepage}
For FI, we use a locally stationary Gaussian thermal covariance at $t_F$, with $q_F=2\Gamma_0T_F$ and $\kappa_{k,F}=G_Fk^2+m_{{\rm eff},F}^2$:
\begin{equation}
\begin{aligned}
\Sigma_{k,11}(t_F)&=\frac{q_F}{2\gamma_{{\rm lin},F}\kappa_{k,F}},\\
\Sigma_{k,22}(t_F)&=\frac{q_F}{2C_F\gamma_{{\rm lin},F}},\qquad
\Sigma_{k,12}(t_F)=0.
\end{aligned}
\label{eq:initial_covariance}
\end{equation}
\end{samepage}
We then set $u_2(t_F)=\dot u_2(t_F)=0$ and include subsequent noise injection. The same state is used in both the two- and three-point functions, including the initial time-current endpoint in Eq.~\eqref{eq:finiteIBP}.

\subsection{Equilateral comparison and independent checks}
\begin{table}[t]
\caption{Same-background equilateral comparison using a common terminal thermal-power normalization. Along this trajectory, the FI unequal-time covariance remains close to its terminal value across much of the interval, whereas FO assumes the decay $e^{-H(t-s)}$. FI also retains the full retarded response. Both calculations utilize the same cubic source, comoving mode, and terminal curvature conversion.}
\label{tab:FO_FI_match}
\centering
\small
\begin{tabular}{rrrr}
\toprule
$N_F$ & \multicolumn{1}{c}{$f_{\rm NL,eq}^{\rm FO}$}
& \multicolumn{1}{c}{$f_{\rm NL,eq}^{\rm FI}$}
& \multicolumn{1}{c}{FO/FI}\\
\midrule
20 & 1.2815 & -417.33 & -0.00307\\
30 & 1.2844 & -371.14 & -0.00346\\
40 & 1.2880 & -326.58 & -0.00394\\
\bottomrule
\end{tabular}

\end{table}
Table~\ref{tab:FO_FI_match} presents the results. At $N_F=30$, $\Xi=0.33328$ and $L_F=4.5197$. The local FO temporal and gradient contributions are $+1.4128$ and $-0.12847$, respectively, whereas the FI kinetic and $K$-derivative contributions are $-375.49$ and $+4.3484$, respectively. The sign reversal and amplitude difference show that the local FO closure fails to provide a quantitative estimate on this trajectory; thus, it is essential to retain unequal-time correlations and the full retarded response.

The Gaussian state is specified as part of the reference prescription. Under the condition of zero initial covariance and the same later thermal noise, FI yields $f_{\rm NL,eq}\simeq-143.17$ at $N_F=30$. For the stationary initial state in the table, $\mathcal P_\zeta^{\rm th}=1.956\times10^{-10}$. These are finite-time thermal-sector quantities instead of a total scalar spectrum or CMB-template fit.

Direct integration of the local current agrees with the closed FO expression at relative differences below $10^{-12}$, and the result is invariant under a joint rescaling of comoving momenta and scale factor. For FI, a finer grid and tighter ODE tolerance lead to $f_{\rm NL,eq}\simeq-371.14$ at $N_F=30$. The independently implemented direct-source and integrated-by-parts forms are consistent in terms of their kinetic and $K$-derivative parts at relative differences below $7.1\times10^{-11}$ and $2\times10^{-14}$, respectively.

Additional exact algebraic checks independently validate the fixed-FLRW quadratic and cubic actions, nonlinear currents and Wick multiplicities, the ADM expansion, and the $\delta N$ sign. The covariance evolution below matches the retarded noise convolution in Eq.~\eqref{eq:Cnoise}; meanwhile, the constant-coefficient Bessel response and the special-function primitives in Appendix~\ref{app:ODreference} have also been examined independently. The verification scripts rely on NumPy and SciPy and use no external data. All comparison assumptions, input values, and reference results are presented in this appendix.

\noindent\textit{AI-assisted computational tools.} DeepSeek-V4 Pro was used to assist with algebraic consistency checks and numerical-code development. The authors specified the physical assumptions, directed the required tasks, independently verified all analytical and numerical results, and take full responsibility for the manuscript.

\subsection{Covariance evolution and analytic response}

\noindent\textit{Joint covariance evolution.}
On an evolving background, the same finite-inertia response determines the stochastic field power spectrum and bispectrum. For each comoving mode, let $\mathbf y_k=(u_k,\dot u_k)^T$ and $\kappa_k=Gk^2/a^2+m_{\rm eff}^2$. The drift matrix and covariance satisfy
\begin{equation}
\begin{gathered}
\mathsf A_k=\begin{pmatrix}0&1\\-\kappa_k/C&-\gamma_{\rm lin}/C\end{pmatrix},\qquad
q=2\Gamma_0Ta^{-3},\\
\dot{\boldsymbol{\Sigma}}_k=\mathsf A_k\boldsymbol{\Sigma}_k+\boldsymbol{\Sigma}_k\mathsf A_k^T
+\frac{q}{C^2}\begin{pmatrix}0&0\\0&1\end{pmatrix}.
\end{gathered}
\label{eq:joint_Lyapunov}
\end{equation}
Let $\mathsf U_k(t,s)$ be the homogeneous fundamental matrix. The covariance and response used in Eq.~\eqref{eq:finiteIBP} are
\begin{equation}
\begin{aligned}
\mathscr C_k(t,s)&=[\mathsf U_k(t,s)\boldsymbol{\Sigma}_k(s)]_{11},\\
\partial_s\mathscr C_k(t,s)&=[\mathsf U_k(t,s)\boldsymbol{\Sigma}_k(s)]_{12},\\
R_k(t,s)&=\mathsf U_{k,12}(t,s)/C(s).
\end{aligned}
\label{eq:joint_CM}
\end{equation}
The Gaussian initial-state choice and the common power/bispectrum normalization are described in Sec.~\ref{subsec:intrinsic}. These equations account for both the propagated initial covariance and the subsequent noise injection.

\noindent\textit{Analytic constant-coefficient response.}
In the de Sitter reference limit with constants $H$, $C$, $G$, $\gamma_{\rm lin}$, and $m_{\rm eff}^2$, define
\begin{equation}
x=\sqrt{G/C}\,\frac{k_c}{aH},\qquad
\nu=\frac{\gamma_{\rm lin}}{2CH},\qquad
\omega_\nu^2=\nu^2-\frac{m_{\rm eff}^2}{CH^2}.
\end{equation}
Here, $k_c$ denotes a comoving wavenumber. The two independent homogeneous mode functions are
\begin{equation}
y_1(t)=x^\nu J_{\omega_\nu}(x),\qquad
y_2(t)=x^\nu Y_{\omega_\nu}(x),
\label{eq:besselmode}
\end{equation}
for which the retarded Green function is
\begin{equation}
R_k(t,s)=\theta(t-s)
\frac{y_1(s)y_2(t)-y_2(s)y_1(t)}
{C\,[y_1(s)\dot y_2(s)-\dot y_1(s)y_2(s)]}.
\label{eq:besselret}
\end{equation}
The Bessel response accounts for inertia, a constant effective mass, and the redshift of physical momentum. Substituting it into Eq.~\eqref{eq:finiteIBP} yields the corresponding finite-inertia field bispectrum. Appendix~\ref{app:ODreference} provides a special-function series for the overdamped reference response.

\noindent\textit{Finite inertia and stochastic contacts.}
With $\gamma_d=3HC+\Gamma_0$, the ratio $\Xi=CH/\gamma_d$ compares inertia over a Hubble time with damping. In a massless, instantaneously fixed-physical-momentum relaxation diagnostic, we have $k_{F,{\rm in}}^2/k_{F,{\rm OD}}^2=1-\Xi$, where $Gk_{F,{\rm OD}}^2=\gamma_d H$.

The finite-inertia Langevin equation also determines a well-defined equal-time velocity variance. In a stationary reference model with a positive restoring coefficient and with $a$, $T$, and the other coefficients held fixed, we have
\begin{equation}
\langle\dot u_1^2\rangle=\frac{\Gamma_0Ta^{-3}}{C\gamma_d}.
\label{eq:stationary_velocity}
\end{equation}
This finite velocity variance complements the endpoint conditions used in Eq.~\eqref{eq:finiteIBP}.

\section{Coupled quadratic action and high-frequency stability}
\label{app:coupled}

For the coupled calculation, we select a minimal scalar-fluid model $P_r(Y)=\lambda Y^2$, where $Y=-g^{\mu\nu}\partial_\mu\sigma\partial_\nu\sigma/2$ and $\lambda>0$. This model gives $p_r=\lambda Y^2$, $\rho_r=3\lambda Y^2$, $c_r^2=1/3$, and $\rho_r+p_r=2Y P_{r,Y}$~\cite{Kase:2014}. For definiteness, a covariant phenomenological energy-transfer choice is specified as
$\nabla_\mu T_{(r)}^{\mu\nu}=Q^\nu$ with
$Q_\mu=-\Gamma(u_r^\nu\nabla_\nu\phi)\nabla_\mu\phi$;
this reproduces $\dot\rho_r+4H\rho_r=\Gamma\dot\phi^2$ in the homogeneous frame. We assume that neither the energy-transfer nor the noise kernels introduce higher-derivative terms to the conservative principal symbol. Under this assumption, the quadratic action below determines the high-frequency stability conditions for the chosen closure.

The Horndeski quantities specialized to the gravity--inflaton sector are
\begin{align}
\Theta&=HM_p^2(1-3Z),\nonumber\\
\Sigma_{\rm H}&=X\left(K+\frac{6X}{\mu^4}\right)
-3H^2M_p^2+\frac{18H^2X}{M^2}.
\label{eq:thetaSigma}
\end{align}
In the scalar-only reference theory, the quadratic curvature coefficients are given by
\begin{equation}
\mathcal G_S=3\mathcal G_T+
\frac{\Sigma_{\rm H}\mathcal G_T^2}{\Theta^2},\qquad
\mathcal F_S=\frac1a\frac{d}{dt}
\left(\frac{a\mathcal G_T^2}{\Theta}\right)-\mathcal F_T.
\label{eq:GSFS}
\end{equation}

We work in the uniform-inflaton gauge, with $N=1+\alpha$, $N_i=\partial_i\chi$, $h_{ij}=a^2e^{2\zeta_u}\delta_{ij}$, and $\sigma=\sigma_0+\pi$. Define $s\equiv\dot\sigma_0$, $D_r\equiv P_{r,Y}+2Y P_{r,YY}$, $b\equiv a^{-2}\partial^2\chi$, and $Z=X/(M_p^2M^2)$. Up to quadratic order, the shift-dependent terms in the scalar--gravity plus fluid action are defined as follows
\begin{equation}
S^{(2)}_{\chi}=\int dt\,d^3x\,a^3
\left[2\{\mathcal G_T\dot\zeta_u-\Theta\alpha\}
+sP_{r,Y}\pi\right]b .
\label{eq:shift2}
\end{equation}
Varying $\chi$ yields Eq.~\eqref{eq:lapseFluid}, or
\begin{equation}
\alpha=A\dot\zeta_u+B\pi,\qquad
A\equiv\frac{\mathcal G_T}{\Theta},\quad
B\equiv\frac{sP_{r,Y}}{2\Theta}.
\label{eq:AB}
\end{equation}
The remaining quadratic kinetic terms before this substitution are given by
\begin{equation}
\begin{aligned}
\frac{\mathcal L_{\mathrm{kin}}^{(2)}}{a^3}={}&
-3\mathcal G_T\dot\zeta_u^2
+6\Theta\alpha\dot\zeta_u
+\Sigma_{\rm H}\alpha^2\\
&+\frac{D_r}{2}(\dot\pi-s\alpha)^2
+\text{lower-derivative terms}.
\end{aligned}
\label{eq:kinBefore}
\end{equation}
We have used the background identity $\Sigma_{\rm H}=p_2-3M_p^2H^2(1-6Z)$, where $p_2=KX+6X^2/\mu^4$.

After substituting Eq.~\eqref{eq:AB} and integrating by parts, the principal part of the quadratic action can be written as
\begin{equation}
\begin{aligned}
S^{(2)}_{\mathrm{prin}}={}&\frac12\int dt\,d^3x\,a^3
\left[\mathsf K_{ab}\dot q_a\dot q_b\right.\\
&\left.\qquad-\frac1{a^2}\mathsf G_{ab}\partial_iq_a\partial_iq_b\right],\\
&(q_1,q_2)=(\zeta_u,\pi),
\end{aligned}
\label{eq:coupledAction}
\end{equation}
where
\begin{equation}
\mathsf K=\begin{pmatrix}
2\mathcal G_S+s^2D_rA^2&-sD_rA\\
-sD_rA&D_r
\end{pmatrix},
\label{eq:KGmatrices}
\end{equation}
\begin{equation}
\mathsf G=\begin{pmatrix}
2\mathcal F_S&-sP_{r,Y}A\\
-sP_{r,Y}A&P_{r,Y}
\end{pmatrix}.
\end{equation}
The determinants factorize exactly,
\begin{align}
\det\mathsf K&=2D_r\mathcal G_S,\\
\det(\mathsf G-c^2\mathsf K)&=
(D_rc^2-P_{r,Y})\nonumber\\[-1mm]
&\quad\times
\bigl[2\mathcal G_Sc^2-2\mathcal F_S
+s^2P_{r,Y}A^2\bigr].
\label{eq:dispersion}
\end{align}
Thus, the two characteristic speeds are given by $c_r^2=P_{r,Y}/D_r=1/3$ and Eq.~\eqref{eq:coupledcs}. Positivity of the kinetic matrix is equivalent to $D_r>0$ and $\mathcal G_S>0$, while the positivity of both squared characteristic speeds additionally entails $P_{r,Y}>0$ and Eq.~\eqref{eq:conditionalstability}. Together, these conditions are necessary and sufficient for the principal quadratic form to be stable in this conservative closure.

For direct application, let $W=X/(H^2M_p^2)$. The Horndeski coefficients reduce to
\begin{equation}
\begin{aligned}
\frac{\mathcal G_S}{M_p^2}=
\frac{1-Z}{(1-3Z)^2}\Biggl[
&W\left(K+\frac{6X}{\mu^4}\right)(1-Z)\\
&+3Z(1+3Z)\Biggr].
\end{aligned}
\end{equation}
\begin{equation}
\begin{aligned}
\frac{\mathcal F_S}{M_p^2}={}&
\frac{(1-Z)^2}{1-3Z}(1+\epsilon)
+\frac{(1-Z)(1+3Z)}{(1-3Z)^2}\frac{\dot Z}{H}\\
&-(1+Z).
\end{aligned}
\end{equation}
In the strict instantaneous fixed-point check, $Z_f=6F/(9F+\overline K)$ and $W_f=6/(9F+\overline K)$. If $\epsilon=\dot Z=0$ and the radiation background is quasi-stationary, $\rho_r\simeq3rX/2$, then the \emph{conservative-fluid} gradient numerator is
\begin{equation}
\begin{aligned}
\frac{\mathcal F_S-(\rho_r+p_r)A^2/2}{M_p^2}
&=\frac{6\mathcal N_{\rm fp}}
{(9F+\overline K)(\overline K-9F)^2},\\
\mathcal N_{\rm fp}&\equiv24F^2(\overline K-9F)
-r(\overline K+3F)^2.
\end{aligned}
\label{eq:gradfixed}
\end{equation}
Equation~\eqref{eq:gradfixed} provides an instantaneous fixed-point diagnostic of the fluid closure. Along an evolving trajectory, the stability domain is determined by the constrained coefficients $\mathcal G_S$ and $\mathcal F_S-(\rho_r+p_r)A^2/2$, as evaluated in Appendix~\ref{app:validation}.

\section{Complete local gravity--inflaton cubic expansion}
\label{app:cubic}

We expand the local Einstein--Hilbert, NMDC, and $P(\phi,X)$ actions to third order in scalar perturbations, working in uniform-inflaton gauge and discarding total spatial derivatives. The minimal radiation-field vertices are listed at the end of the appendix.

With $N=1+\alpha$, $N_i=\partial_i\chi$, $h_{ij}=a^2e^{2\zeta_u}\delta_{ij}$, introduce
\begin{equation}
 B_{ij}=a^{-2}\partial_i\partial_j\chi,
 \quad b=B_{ii},\quad d=a^{-2}\partial_i\zeta_u\partial_i\chi,
\end{equation}
\begin{equation}
 Q_B=B_{ij}B_{ij}-b^2,\quad
 J=a^{-2}B_{ij}\partial_i\zeta_u\partial_j\chi.
\end{equation}
The exact mixed extrinsic curvature is
\begin{equation}
\begin{aligned}
K^i{}_j=\frac1N\Bigl[(H+\dot\zeta_u)\delta^i_j
-a^{-2}e^{-2\zeta_u}\bigl(&\partial^i\partial_j\chi\\
&-\partial^i\zeta_u\partial_j\chi
-\partial_j\zeta_u\partial^i\chi\\
&+\delta^i_j\partial_k\zeta_u\partial_k\chi\bigr)\Bigr].
\end{aligned}
\end{equation}
Let
\begin{align}
c_0&=1-Z,&c_1&=-1+3Z,\\
c_2&=1-6Z,&c_3&=-1+10Z,
\end{align}
and define the perturbative coefficients
\begin{align}
 D_0={}&c_0,\nonumber\\
 D_1={}&c_1\alpha+3c_0\zeta_u,\nonumber\\
 D_2={}&c_2\alpha^2+3c_1\alpha\zeta_u
 +\tfrac92c_0\zeta_u^2,\nonumber\\
 D_3={}&c_3\alpha^3+3c_2\alpha^2\zeta_u
 +\tfrac92c_1\alpha\zeta_u^2
 +\tfrac92c_0\zeta_u^3.
\end{align}
The extrinsic-curvature invariant, with its overall lapse prefactor omitted, has a perturbative expansion $S_0+S_1+S_2+S_3$ with
\begin{align}
 S_0={}&-6H^2,\nonumber\\
 S_1={}&-12H\dot\zeta_u+4Hb,\nonumber\\
 S_2={}&-6\dot\zeta_u^2+4Hd-8H\zeta_u b
 +4\dot\zeta_u b+Q_B,\nonumber\\
 S_3={}&8H\zeta_u^2b-8H\zeta_u d
 +4\dot\zeta_u d-8\dot\zeta_u\zeta_u b\nonumber\\
 &-4\zeta_uQ_B-4J.
\label{eq:Si}
\end{align}
Then, the complete extrinsic contribution at cubic order is given by
\begin{equation}
\mathcal L^{(3)}_{K}/a^3
=\frac{M_p^2}{2}
\left(D_0S_3+D_1S_2+D_2S_1+D_3S_0\right).
\label{eq:LK3}
\end{equation}
For the purely kinetic inflaton, introduce
\begin{align}
 p_0&=KX+\frac{X^2}{\mu^4},&
 p_1&=-KX-3\frac{X^2}{\mu^4},\nonumber\\
 p_2&=KX+6\frac{X^2}{\mu^4},&
 p_3&=-KX-10\frac{X^2}{\mu^4}.
\end{align}
Its complete cubic term is written as
\begin{equation}
\mathcal L^{(3)}_{P}/a^3
=p_3\alpha^3+3p_2\alpha^2\zeta_u
+\tfrac92p_1\alpha\zeta_u^2
+\tfrac92p_0\zeta_u^3.
\label{eq:LP3}
\end{equation}
Finally, the complete spatial-curvature cubic term is shown in Eq.~\eqref{eq:metriccubicR3} and is obtained by expanding
\begin{equation}
\begin{aligned}
S_{R_3}=M_p^2\!\int\!dt\,d^3x\,a e^{\zeta_u}
\Big[&\left(N+\frac{Z}{N}\right)
(\partial\zeta_u)^2\\
&+2\left(1-\frac{Z}{N^2}\right)
\partial_iN\partial_i\zeta_u\Big].
\end{aligned}
\end{equation}
Equations~\eqref{eq:LK3}, \eqref{eq:LP3}, and \eqref{eq:metriccubicR3} account for all \emph{local gravity--inflaton} scalar cubic vertices before solving constraints.

For completeness, one may also include a local minimal radiation-field contribution. Let $s=\dot\sigma_0$, $g=a^{-2}\partial_i\chi\partial_i\pi$, and $h=a^{-2}(\partial\pi)^2$. Expanding $Y=Y_0+Y_1+Y_2+Y_3+\cdots$ yields
\begin{align}
 Y_0={}&s^2/2,\\
 Y_1={}&s\dot\pi-s^2\alpha,\\
 Y_2={}&\tfrac12\dot\pi^2-2s\alpha\dot\pi
 +\tfrac32s^2\alpha^2-sg-\tfrac12h,\\
 Y_3={}&-\alpha\dot\pi^2+3s\alpha^2\dot\pi-2s^2\alpha^3
 +2s\alpha g-\dot\pi g\nonumber\\
 &+2s\zeta_u g+\zeta_u h.
\end{align}
The cubic term of $N\sqrt h\,P_r(Y)$ is given by
\begin{equation}
\begin{aligned}
\mathcal L_r^{(3)}/a^3={}&
\frac92P_r\zeta_u^2(\zeta_u+\alpha)\\
&+P_{r,Y}\bigl[Y_3+(\alpha+3\zeta_u)Y_2\\
&\qquad +(3\alpha\zeta_u+\tfrac92\zeta_u^2)Y_1\bigr]\\
&+P_{r,YY}\bigl[Y_1Y_2
+\tfrac12(\alpha+3\zeta_u)Y_1^2\bigr]\\
&+\frac16P_{r,YYY}Y_1^3.
\end{aligned}
\label{eq:fluidcubic}
\end{equation}
where all $P_r$ derivatives are evaluated at $Y_0$. When $P_r=\lambda Y^2$, the last term vanishes. Equations~\eqref{eq:LK3}--\eqref{eq:fluidcubic} define the local gravity--inflaton and minimal fluid vertices. A full metric--radiation curvature bispectrum further involves dissipative transfer interactions, which depend on the underlying matter model.

\section{Auxiliary overdamped reference response}
\label{app:ODreference}
We derive a special-function series for an overdamped reference response, assuming that $H,C,G,\Gamma_0,T$ are positive constants over $[t_F,t]$, neglecting $m_{\rm eff}^2$, and writing $\gamma_d=3HC+\Gamma_0$. The resulting linear response approximates the inertial theory when $CH/\gamma_d\ll1$; this condition does not hold on the near-attractor branch of Eq.~\eqref{eq:chi_branch}. Since an ideal first-order white-noise process has no ordinary velocity variance, the following derivative vertices are defined through separated-time contractions and one-sided covariance derivatives. Taking $a(t_F)=1$, define
\begin{equation}
\begin{gathered}
y=e^{-2H(s-t_F)},\quad Y=e^{-2H(t-t_F)},\\
\beta_i=\frac{Gk_i^2}{2H\gamma_d},\qquad
R_k^{\rm OD}(t,s)=\frac{e^{-\beta_k(y-Y)}}{\gamma_d}.
\end{gathered}
\label{eq:ODgreen}
\end{equation}
For independent Gaussian initial data of power $\mathscr P_{iF}$, the covariance $\mathscr C_i(t,s)=e^{\beta_i y}E_i(y)$, $t\ge s$, satisfies
\begin{equation}
\begin{aligned}
E_i(y)={}&\mathscr P_{iF}e^{-\beta_i(2-Y)}\\
&+\frac{\Gamma_0T}{H\gamma_d^2}e^{\beta_iY}
\sum_{n=0}^{\infty}\frac{(-2\beta_i)^n(1-y^{n+3/2})}{n!(n+3/2)}.
\end{aligned}
\label{eq:ODcov}
\end{equation}
This series converges absolutely over the finite interval. Define $V_i(y)=-2Hy[\beta_iE_i(y)+E_i'(y)]$, so that $\partial_s\mathscr C_i(t,s)=e^{\beta_i y}V_i(y)$.

We begin by differentiating the time current in Eq.~\eqref{eq:contractedS} and then evaluate the local coefficients at a reference time. With $K_n=d^nK/d\phi^n$, the relations $\dot K_1=K_2v$ and $\dot K_2=K_3v$ preserve the derivatives of the vertex coefficients. Keeping the resulting values of $v$, $\dot v$, and $K_n$ fixed, and defining $P_{ij}=\mathbf k_i\cdot\mathbf k_j$ and $\mathbf k_\ell=-\mathbf k_i-\mathbf k_j$, the pairwise contractions are given by
\begin{equation}
\begin{aligned}
J_{ij}={}&\frac{2v}{\mu^4}
 [3V_iV_j+P_{ij}yE_iE_j]\\
&+K_{\phi\phi}vE_iE_j
 +K_\phi(V_iE_j+E_iV_j),
\end{aligned}\label{eq:allJ}
\end{equation}
\begin{equation}
\begin{aligned}
P_{ij}^{(\phi)}={}&K_{\phi\phi\phi}X_0E_iE_j
 +K_{\phi\phi}v(V_iE_j+E_iV_j)\\
&+K_\phi[V_iV_j+P_{ij}yE_iE_j].
\end{aligned}\label{eq:allP}
\end{equation}
The contracted spatial-current divergence is denoted by $yW_{ij|\ell}$, where
\begin{equation}
\begin{aligned}
W_{ij|\ell}={}&\frac{2v}{\mu^4}
 [(\mathbf k_\ell\cdot\mathbf k_j)V_iE_j\\
 &\hspace{11mm}+(\mathbf k_\ell\cdot\mathbf k_i)E_iV_j]
 -K_\phi k_\ell^2E_iE_j.
\end{aligned}\label{eq:allW}
\end{equation}
When the following primes act only on the explicit $y$ dependence of $E_i,V_i$ and $y$, the derivatives of the background coefficients provide an additional term,
\begin{equation}
\begin{aligned}
\Delta\mathcal S_{ij}={}&-(K_3v^2+K_2\dot v)E_iE_j\\
&-K_2v(V_iE_j+E_iV_j)\\
&-\frac{6\dot v}{\mu^4}V_iV_j
-\frac{2\dot vP_{ij}}{\mu^4}yE_iE_j.
\end{aligned}
\label{eq:vertex_coefficient_derivative}
\end{equation}
As a local check, the net $K_3$ contribution at $K_1=K_2=\dot v=0$ is $-K_3X_0E_iE_j$. The reference contracted source is represented as $e^{(\beta_i+\beta_j)y}\mathcal S^{\rm all}_{ij|\ell}(y)$, with
\begin{equation}
\begin{aligned}
\mathcal S^{\rm all}_{ij|\ell}(y)={}&
2Hy[J'_{ij}+(\beta_i+\beta_j)J_{ij}]\\
&-3HJ_{ij}+P_{ij}^{(\phi)}+yW_{ij|\ell}\\
&+\Delta\mathcal S_{ij}\\
={}&\sum_{p\ge0}s^{\rm all}_{ij|\ell,p}y^p.
\end{aligned}\label{eq:fullKpoly}
\end{equation}
All coefficient functions are specified explicitly in Eqs.~\eqref{eq:ODcov}, \eqref{eq:allJ}--\eqref{eq:fullKpoly}; equivalently, we have $s^{\rm all}_{ij|\ell,p}=[y^p]\mathcal S^{\rm all}_{ij|\ell}(y)$. Generalized powers $p=0,1,3/2,\ldots$ occur. In particular, the $K$-derivative vertices contribute a constant term, which does not appear in the kinetic-only calculation. For $p>0$, the time primitive is
\begin{equation}
\begin{aligned}
\mathcal M_p(A,Y)={}&\frac1p\bigl[{}_1F_1(p;p+1;A)\\
&-Y^p{}_1F_1(p;p+1;AY)\bigr],\quad p>0.
\end{aligned}
\label{eq:Mp}
\end{equation}
The $p=0$ primitive is
\begin{equation}
\mathcal M_0(A,Y)=
\begin{cases}
\operatorname{Ei}(A)-\operatorname{Ei}(AY),& A\ne0,\\
-\ln Y,& A=0.
\end{cases}\label{eq:M0}
\end{equation}
The resulting convergent reference field-bispectrum series, which retains the local coefficients of every fixed-metric scalar cubic vertex, is
\begin{equation}
\begin{aligned}
B_u^{\rm all,OD}={}&\sum_{\rm cyc(ij|\ell)}
\frac{e^{\beta_\ell Y}}{2H\gamma_d}
\sum_{p\ge0}s^{\rm all}_{ij|\ell,p}
\mathcal M_p(A_{ij|\ell},Y),
\end{aligned}\label{eq:closed_all_K}
\end{equation}
where $A_{ij|\ell}=\beta_i+\beta_j-\beta_\ell$. Here, $\mathscr P_i(t)=e^{\beta_iY}E_i(Y)$. Within the curvature sign convention adopted above, we have
\begin{equation}
f_{\rm NL}^{\rm all,OD}=-\frac{5v}{6H}
\frac{B_u^{\rm all,OD}}{\sum_{\rm cyc}\mathscr P_i(t)\mathscr P_j(t)}.
\label{eq:closed_all_fnL}
\end{equation}
For equal momenta, one has $P_{ij}=-k^2/2$ and $\beta_i=\beta$. Substituting these relations directly into Eq.~\eqref{eq:fullKpoly} yields $s_{\rm eq,p}$, and then the result simplifies to
\begin{equation}
f_{\rm NL,eq}^{\rm all,OD}=
-\frac{5v}{12H^2\gamma_d}\frac{e^{-\beta Y}}{E(Y)^2}
\sum_{p\ge0}s_{\rm eq,p}\mathcal M_p(\beta,Y).
\label{eq:equilateral_all_K}
\end{equation}
The functions $\mathcal M_p$ encode the distinct response weights associated with the cubic interactions. To match this reference contraction to a physical nonlinear overdamped stochastic model, the noise regularization scheme must be specified.

%------------------------------------------------------------------------------------------------------------------------

\end{document}